\documentclass[aps,prl,reprint, showpacs,floatfix,amsmath,amssymb,superscriptaddress]{revtex4-1}
\usepackage{graphicx}
\begin{document}

\title{Synthesis of high-oxidation Y-Ba-Cu-O phases in superoxygenated thin films}

\author{H. Zhang}
\affiliation{Department of Physics, University of Toronto, Toronto, M5S1A7, Canada}

\author{N. Gauquelin}
\thanks{Current Address: EMAT, Department of Physics, University of Antwerp, Groenenborgerlaan 171, 2020 Antwerp, Belgium}
\affiliation{Department of Materials Science and Engineering, McMaster University, 1280 Main Street West, Hamilton, Ontario, L9H4L7, Canada}

\author{C. McMahon}
\affiliation{Department of Physics and Astronomy, University of Waterloo, Waterloo, N2L3G1, Canada}

\author{D. G. Hawthorn}
\affiliation{Department of Physics and Astronomy, University of Waterloo, Waterloo, N2L3G1, Canada}
\affiliation{Canadian Institute for Advanced Research, Toronto, M5G1Z8, Canada}

\author{G. A. Botton}
\affiliation{Department of Materials Science and Engineering, McMaster University, 1280 Main Street West, Hamilton, Ontario, L9H4L7, Canada}

\author{J. Y.T. Wei}
\affiliation{Department of Physics, University of Toronto, Toronto, M5S1A7, Canada}
\affiliation{Canadian Institute for Advanced Research, Toronto, M5G1Z8, Canada}
\begin{abstract}

It is known that solid-state reaction in high-pressure oxygen can stabilize high-oxidation phases of Y-Ba-Cu-O superconductors in powder form.  We extend this superoxygenation concept of synthesis to thin films which, due to their large surface-to-volume ratio, are more reactive thermodynamically.  Epitaxial thin films of $\rm{YBa_2Cu_3O_{7-\delta}}$ grown by pulsed laser deposition are annealed at up to 700 atm O$_2$ and 900$^\circ$C, in conjunction with Cu enrichment by solid-state diffusion.  The films show clear formation of $\rm{Y_2Ba_4Cu_7O_{15-\delta}}$ and $\rm{Y_2Ba_4Cu_8O_{16}}$ as well as regions of $\rm{YBa_2Cu_5O_{9-\delta}}$ and YBa$_2$Cu$_6$O$_{10-\delta}$ phases, according to scanning transmission electron microscopy, x-ray diffraction and x-ray absorption spectroscopy.  Similarly annealed $\rm{YBa_2Cu_3O_{7-\delta}}$ powders show no phase conversion. Our results demonstrate a novel route of synthesis towards discovering more complex phases of cuprates and other superconducting oxides. 

\end{abstract}

\maketitle

\section{I. INTRODUCTION}
The superconducting critical temperature $T_c$ of hole-doped cuprate tends to scale with their lattice complexity, which is generally correlated with increasing oxidation of the cation block \cite{chu, bianconi, sarrao}. For the Y-Ba-Cu-O family, best known for $\rm{YBa_2Cu_3O_{7-\delta}}$ (YBCO-123) \cite{MKWu, cava}, earlier studies have shown that solid-state reaction of powder samples in high O$_2$ pressure can stabilize the formation of higher-oxidation phases such as $\rm{Y_2Ba_4Cu_7O_{15-\delta}}$ (YBCO-247) and $\rm{Y_2Ba_4Cu_8O_{16}}$ (YBCO-248) \cite{bordet,karpinski248, karpinski,morris, morris2}. These phases are distinguished by double-CuO chains that either replace or alternate with single-CuO chains, resulting in longer unit cells along the $c$-axis \cite{mandich, marsh, cava248, tallon}. More complex phases, with higher ratios of CuO chains to CuO$_2$ planes, are also believed to exist but have not yet been synthesized, likely due to limitations in thermodynamic stability \cite{ramesh, ramesh2, ramesh3, senaris, voronin}.  Oxidation states of the various Y-Ba-Cu-O phases are listed in Table 1. The total cation valence is defined as the sum of valences for all the cations in one block of $\rm{\frac{1}{2}(YBa_2Cu_{3+x/2}O_{7+x/2})}$ that contains one CuO$_2$ plane, and scales with the ratio of CuO chains to CuO$_2$ planes.

In this study, we extend this superoxygenation concept of oxide synthesis to Y-Ba-Cu-O thin films which, due to their large surface-to-volume ratio, are thermodynamically more reactive than powders or crystals \cite{naito}. Epitaxial thin films of YBCO-123 are grown on lattice-matched perovskite substrates by pulsed laser-ablated deposition (PLD), and then post-annealed in high O$_2$ pressure both at and above the PLD growth temperature. The high-pressure annealing is done in conjunction with Cu enrichment by solid-state diffusion, to facilitate the inclusion of extra CuO chains. Scanning transmission electron microscopy (STEM), x-ray diffraction (XRD) and x-ray absorption spectroscopy (XAS) are used to characterize the films.  The post-annealed films show clear formation of YBCO-247 and YBCO-248 phases, as well as regions of $\rm{YBa_2Cu_5O_{9-\delta}}$ (YBCO-125) and $\rm{YBa_2Cu_6O_{10-\delta}}$ (YBCO-126) phases having triple-CuO and quadruple-CuO chains, respectively.  Similarly annealed YBCO-123 powders show no phase conversion. These observations demonstrate that high-oxidation phases of Y-Ba-Cu-O can be stabilized in thin-film form by annealing in high-pressure oxygen. As proof of concept, our results open a novel route of synthesis for discovering more complex phases of high-$T_c$ cuprates and other superconducting oxides. 

\begin{table}[htb]
\centering
\begin{tabular}{|l|l|l|l|l|}
\hline
Formula Unit & Y:Ba:Cu & Total Cation Valence & CuO/CuO$_2$\\ \hline
YBa$_2$Cu$_3$O$_{7}$   & 123  &   +7      & 0.5                        \\ \hline
Y$_2$Ba$_4$Cu$_7$O$_{15}$   & 247  &          +7.5    & 0.75           \\ \hline
Y$_2$Ba$_4$Cu$_8$O$_{16}$   & 248  &          +8      &  1                \\ \hline
Y$_2$Ba$_4$Cu$_9$O$_{17}$   & 249  &     +8.5   & 1.25             \\ \hline
YBa$_2$Cu$_5$O$_{9}$   & 125         &      +9      &  1.5                  \\ \hline
YBa$_2$Cu$_6$O$_{10}$ & 126         &      +10      &  2                  \\ \hline
\end{tabular}
\caption{Known and possible phases of the Y-Ba-Cu-O family of cuprates, listed by nominal formula unit and ratio of cations.  The total cation valence (relative to one CuO$_2$) scales with the ratio of CuO chains to CuO$_2$ planes. For simplicity, oxygen non-stoichiometry is not shown.}
\end{table}
\section{II. EXPERIMENTAL}

The PLD apparatus used for our experiment is equipped with a KrF excimer laser operating at 248 nm, 2 Hz and 2 $\rm{J/cm^2}$. The ceramic YBCO-123 target used is $>99.9\%$ pure and $\sim93\%$ dense.  Thin films of YBCO-123 were grown on $c$-axis faces of $\rm{(LaAlO_3)_{0.3}(Sr_2TaAlO_6)_{0.7}}$ (LSAT) substrates, at 800 $^\circ$C in 200 mTorr of $\rm{O_2}$.  The films were 50 nm thick and 2 nm smooth, as determined by STEM and atomic force microscopy.  After deposition, each film was annealed \textit{in-situ} by cooling at 12$^\circ$C/min to 300$^\circ$C in 760 Torr of $\rm{O_2}$ to ensure proper oxygenation of YBCO-123. The resistively-measured $T_c$ of our as-grown films are $\sim$ 91$\pm$1 K, comparable to optimally-doped YBCO-123 single crystals. The as-grown films were post-annealed with a commercial high-pressure (HP) furnace, at up to 700 atm O$_2$ and 900$^\circ$C, for 12 hours before being cooled in the HP oxygen.  During the HP-annealing, each film was buried under YBCO-123 powder enriched with CuO powder, to enable solid-phase diffusion of the excess Cu.  For the enrichment, CuO to YBCO-123 molar ratios of 2:1 and 4:1 were used, yielding HP-annealed films that are respectively denoted as Samples A and B in this paper. These samples were annealed in 650 atm of O$_2$ at 800$^\circ$C, which we found was the optimal temperature for driving phase conversion without causing decomposition.

XRD measurements were done using the $\theta$ - 2$\theta$ method with a Bruker D8 DISCOVER X-ray diffractometer.  STEM measurements were made with a FEI Titan 80-300 microscope, which is fitted with a high-brightness field emission gun and CEOS aberration correctors for both condenser and objective lens aberrations, and operated at 200 keV in scanning mode.  The resulting high-angle annular dark-field (HAADF) images, that are sensitive to atomic-number contrast. XAS measurements were performed at the Canadian Light Source's REIXS beamline using total fluorescence yield (TFY) mode at room temperature \cite{hawthorn-setup}. Data was taken for both Cu $L$ edge and O $K$ edge on all samples, with the orientation of the linear polarization aligned along both the in-plane (\textbf{E} $\parallel ab$) and the out-of-plane (\textbf{E} $\parallel c$) directions.

\section{III. RESULTS AND DISCUSSIONS}
Figure 1 compares the STEM and XRD data between our HP-annealed and as-grown films, the former showing significant conversion of YBCO-123 to both YBCO-247 and YBCO-248, as represented by the equations:
\begin{equation*}
\begin{split}
  \rm{2YBa_2Cu_3O_{7-\delta} + CuO \xrightarrow{HP\ O_2} Y_2Ba_4Cu_7O_{15-\delta}}\\
  \rm{2YBa_2Cu_3O_{7-\delta} + 2CuO \xrightarrow{HP\ O_2} Y_2Ba_4Cu_8O_{16}}
\end{split}
\end{equation*}

Fig. 1(a) shows the STEM image taken on the HP-annealed Sample A. The double-CuO chains appear as dark stripes and are labelled as D, while the single-CuO chains are labelled as S.  In this image we see a gradation of phases, with YBCO-248 occurring near the surface, followed by YBCO-247 below that and then YBCO-123 near the film-substrate interface. This gradation of phases can be understood in terms of a natural tendency for chain-rich Y-Ba-Cu-O phases with higher cation valence to form closer to the film surface, since it is easier for both  O and Cu ions to diffuse into the top part than into the bottom part of the film. On the other hand, the STEM image of a typical as-grown film in Fig. 1(b) shows pure YBCO-123 containing only single-CuO chains.  Images taken on 1-atm annealed films (data not shown) have similar epitaxy and film purity as those of as-grown films. To corroborate the STEM images, Figs. 1(c) and 1(d) show the XRD patterns for Sample A and the as-grown film respectively. In these figures the peaks associated with the LSAT substrate are not labelled, since the locations of all the substrate peaks coincide with some of the Y-Ba-Cu-O peaks.  For Sample A, all the peaks associated with YBCO-123, YBCO-247 and YBCO-248 are present, indicating a mixture of phases consistent with the STEM image shown in Fig. 1(a). For the as-grown film, on the other hand, all the expected XRD peaks associated with the \textit{c}-axis of YBCO-123 are present, and there are no impurity peaks to within the resolution of our instrument. By relating the YBCO-123 (005)- and (007)-peaks with $2\theta = 38.58^\circ$ and $2\theta = 55.08^\circ$, respectively, we find a \textit{c}-axis lattice parameter of 11.67 \AA\ for our as-grown film, consistent with fully-oxygenated YBCO-123 \cite{capponi, beno}. 
\begin{figure}[t]
\centering
\includegraphics[width=3.35in]{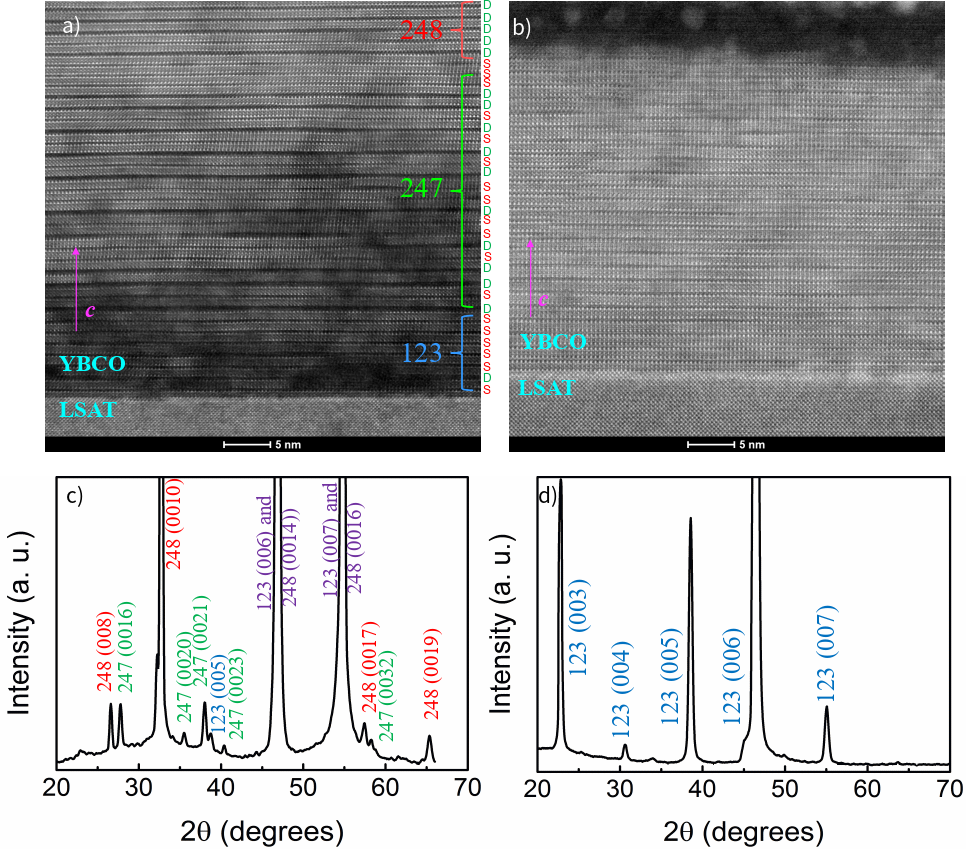}
\caption{Comparison of the STEM and XRD data between HP-annealed and as-grown films. (a) shows STEM image of the HP-annealed Sample A, showing a gradation (bottom to top) of phase conversion from YBCO-123 to YBCO-247 and YBCO-248, with single-CuO chains labelled as S and double-CuO chains labelled as D.  (b) shows STEM image of an as-grown film, showing pure YBCO-123 containing only single-CuO chains.  (c) and (d) show the XRD patterns of sample A and an as-grown film, respectively.  The XRD data are consistent with the phases seen by STEM.}
\end{figure}

\begin{figure}[t]
\centering
\includegraphics[width=3.3in]{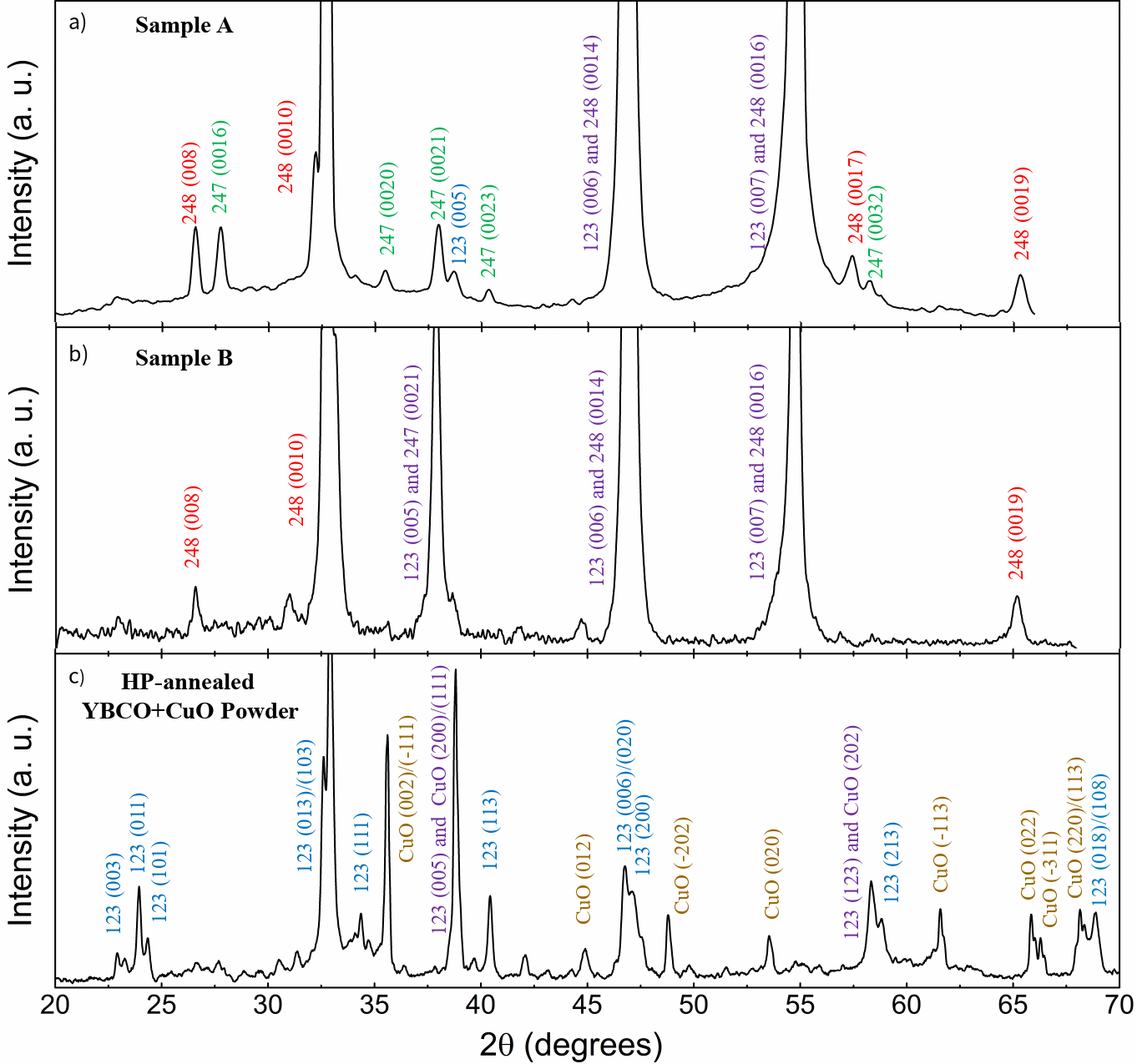}
\caption{Comparison of XRD data between different HP-annealed films, along with a HP-annealed powder serving as control. (a) and (b) show the data for Samples A and B, which were HP-annealed with Cu enrichment using 2:1 and 4:1 molar ratio of CuO to YBCO-123, respectively.  (c) shows the data for YBCO-123 powder that was HP-annealed after being mixed with CuO powder in 4:1 molar ratio.  Whereas higher Cu enrichment promoted conversion to higher-oxidation phases in the thin films, there is no sign of phase conversion in the powder sample.}
\end{figure}

The conversion to higher-oxidation phases induced by HP annealing in our YBCO-123 films is enhanced by higher Cu enrichment. Figure 2 compares the XRD data between different HP-annealed films, as well as HP-annealed powder serving as a control. Figs. 2(a) and 2(b) show the data for Samples A and B, which were HP-annealed with Cu enrichment using 2:1 and 4:1 molar ratio of CuO to YBCO-123, respectively.  In contrast to the mixture of 123, 247 and 248 peaks seen in Sample A, almost all the 247 peaks have disappeared from Sample B, indicating more thorough conversion to the higher-oxidation 248 phase. As a further corroboration, Figure 3 compares the resistance ($R$)  {\it vs.} temperature ($T$) data between Sample A and Sample B.  Sample B shows $T_c$ $\sim$ 80 K, consistent with the predominance of YBCO-248 phase, which is known to have $T_c$ $\sim$ 80 K because of its inherent underdoping \cite{buckley}. Sample A shows $T_c$ $\sim$ 91 K, consistent with a mixture of phases where the significant presence of YBCO-123 provides higher-$T_c$ conduction paths. Finally, we note that there is no sign of phase conversion in powder YBCO-123 samples that were similarly HP-annealed, even for 4 times as long.  As evidenced by the XRD data in Fig. 2(c), YBCO-123 powder that was HP-annealed for 2 days in 650 atm of O$_2$ at 800$^\circ$C, after being mixed with CuO powder in 4:1 molar ratio, shows only YBCO-123 and CuO peaks, just as before the annealing.  This difference between the HP-annealed film versus powder provides strong evidence that the large surface-to-volume ratio of thin films is thermodynamically favoring the phase conversion. 

\begin{figure}[t]
\centering
\includegraphics[width=3.2in]{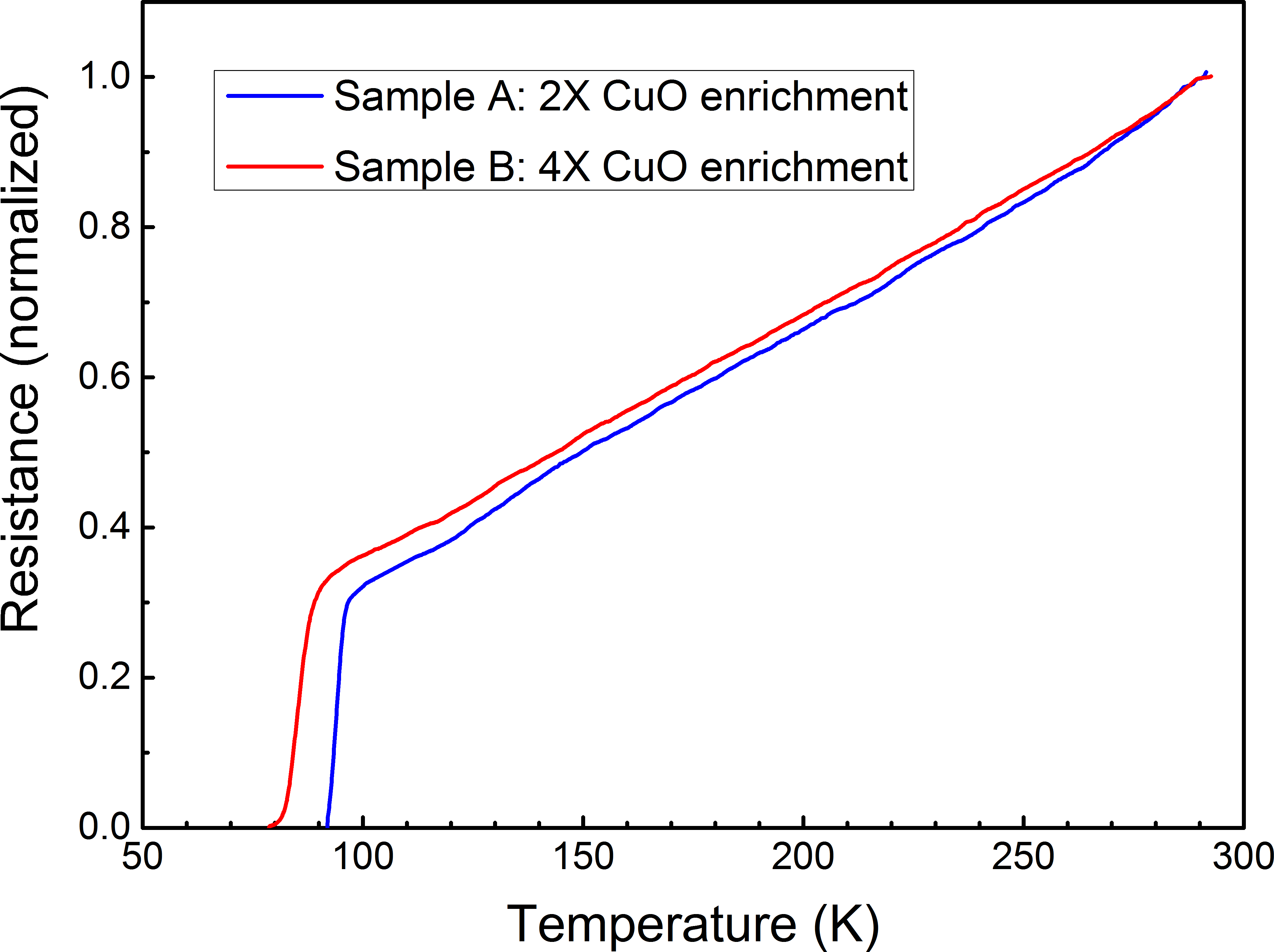}
\caption{Comparison of the resistance {\it vs.} temperature data between Sample A and Sample B, normalized relative to their room-temperature values.  Sample B shows $T_c$ $\sim$ 80 K, consistent with the predominance of YBCO-248 phase as indicated by both the STEM and XRD data.  Sample A shows $T_c$ $\sim$ 91 K, consistent with a mixture of phases where the significant presence of YBCO-123 provides higher-$T_c$ conduction paths.}
\end{figure}

Figure 4 shows the corrected and normalized XAS spectra taken on our as-grown and HP-annealed films. The XAS spectra with \textbf{E} $\parallel ab$ are shifted up for clarity. Panel (a) shows the Cu $L_3$ absorption edge spectra, which arises from the $2p \to 3d$ transition \cite{fink}. In the \textbf{E} $\parallel ab$ spectra, the main peak at 931 eV is mostly associated with the Cu(2) atoms in the planes, and do not show much variation between different samples. Slightly above the main peak in the \textbf{E} $\parallel ab$ spectra, there is a shoulder at 932 eV that is caused by the presence of itinerant holes in the plane \cite{chen, pellegrin, garg, hawthorn}. Surprisingly, our data show that both HP-annealed films have fewer holes in the planes than does the as-grown film, because the shoulder is weakened after annealing. Sample A, which contains both YBCO-247 and YBCO-248 in additional to YBCO-123, appears to be more underdoped in the planes than Sample B, which lacks YBCO-247. Since YBCO-248 is known to be inherently underdoped with little oxygen variability \cite{buckley}, it is reasonable that the as-grown film is less underdoped than any of the HP-annealed films, all of which contain YBCO-248. More interestingly, our data seem to suggest that the film containing YBCO-247 is more underdoped than the films without YBCO-247 because the shoulder at 932 eV is the weakest in Sample A. This observation could explain the generally lower $T_c$ reported in pure YBCO-247 samples \cite{kato, irizawa}. On the other hand, in the \textbf{E} $\parallel c$ data, the peak near 931 eV is mostly associated with the Cu(1) atoms in the CuO chains \cite{nucker}. It is clear that Sample B has the strongest peak, followed by Sample A, and the as-grown film has the weakest peak. This observation is consistent with the natural expectation that higher Cu enrichment introduces more CuO chains into the Y-Ba-Cu-O lattice.

\begin{figure}[t]
\centering
\includegraphics[width=3.2in]{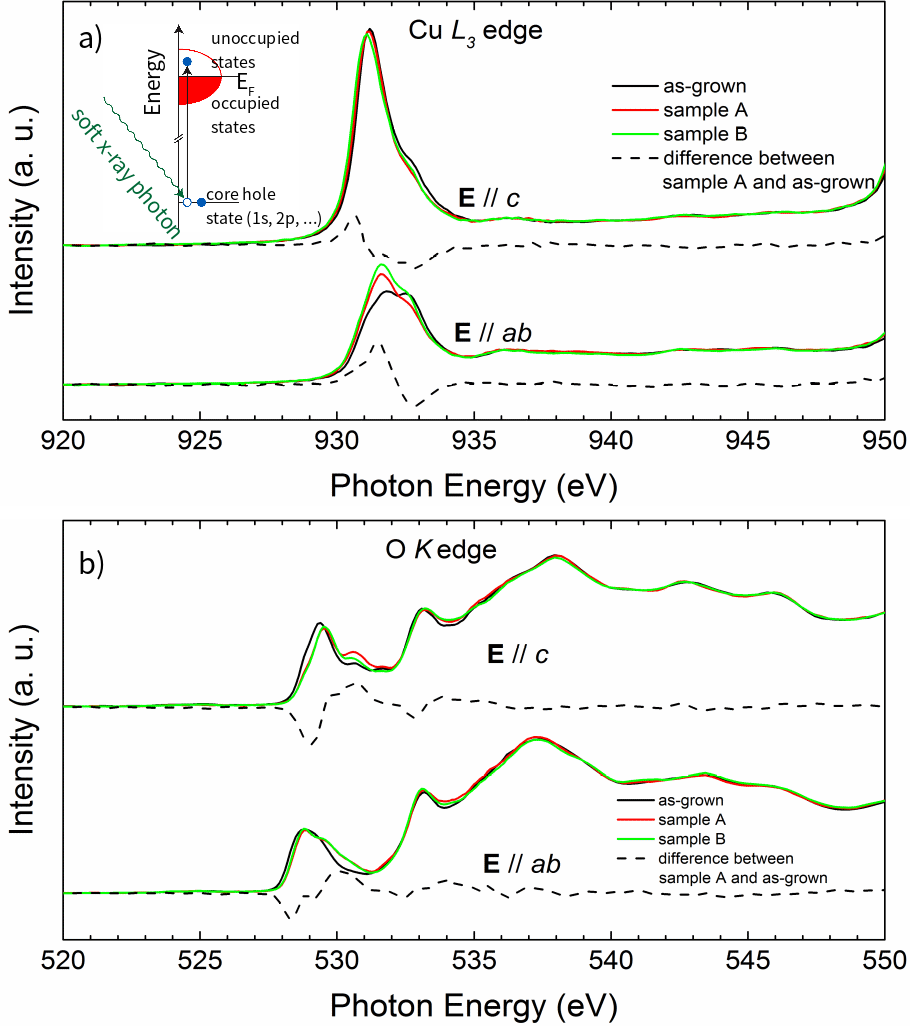}
\caption{Corrected and normalized XAS spectra taken on both as-grown and HP-oxygen annealed films. (a) shows the spectra for the Cu $L_3$ edge, with a schematic diagram of the XAS process in the inset. (b) shows the spectra for the O $K$ edge. The XAS spectra show that there are more chain states in the annealed films. There is also evidence that the $\rm{CuO_2}$ planes are underdoped in the HP-annealed films, with the sample containing YBCO-247 being the most underdoped.}
\end{figure}

The O $K$ edge absorption edge spectra is shown in Fig. 4(b).  The peaks at $\sim 529$ eV and $\sim 530$ eV in the \textbf{E} $\parallel ab$ data represent the Zhang-Rice singlet band (ZRSB) and the upper Hubbard band (UHB) respectively, and a shifting of spectral weight from the ZRSB to the UHB indicates a reduction of holes in the planes \cite{nucker}. The as-grown sample has the weakest UHB spectral weight relative to the ZRSB, while Sample A, which contains large regions of YBCO-247, has the strongest UHB. This observation is again consistent with the XAS data for the Cu $L_3$ absorption edge, which shows that the sample containing pure YBCO-123 is the least underdoped, and that the sample containing a mixture of YBCO-123, YBCO-248 and YBCO-247 is the most underdoped. In the \textbf{E} $\parallel c$ spectrum, the peak at $\sim 529$ eV represents the O $\rm{2p_z}$ states from apical O(4) sites \cite{nucker}, and this peak in both HP-annealed films are broader than in the as-grown film, implying that HP annealing increases the number of holes in the apical sites.

\begin{figure}[hbt]
\centering
\includegraphics[width=3.2in]{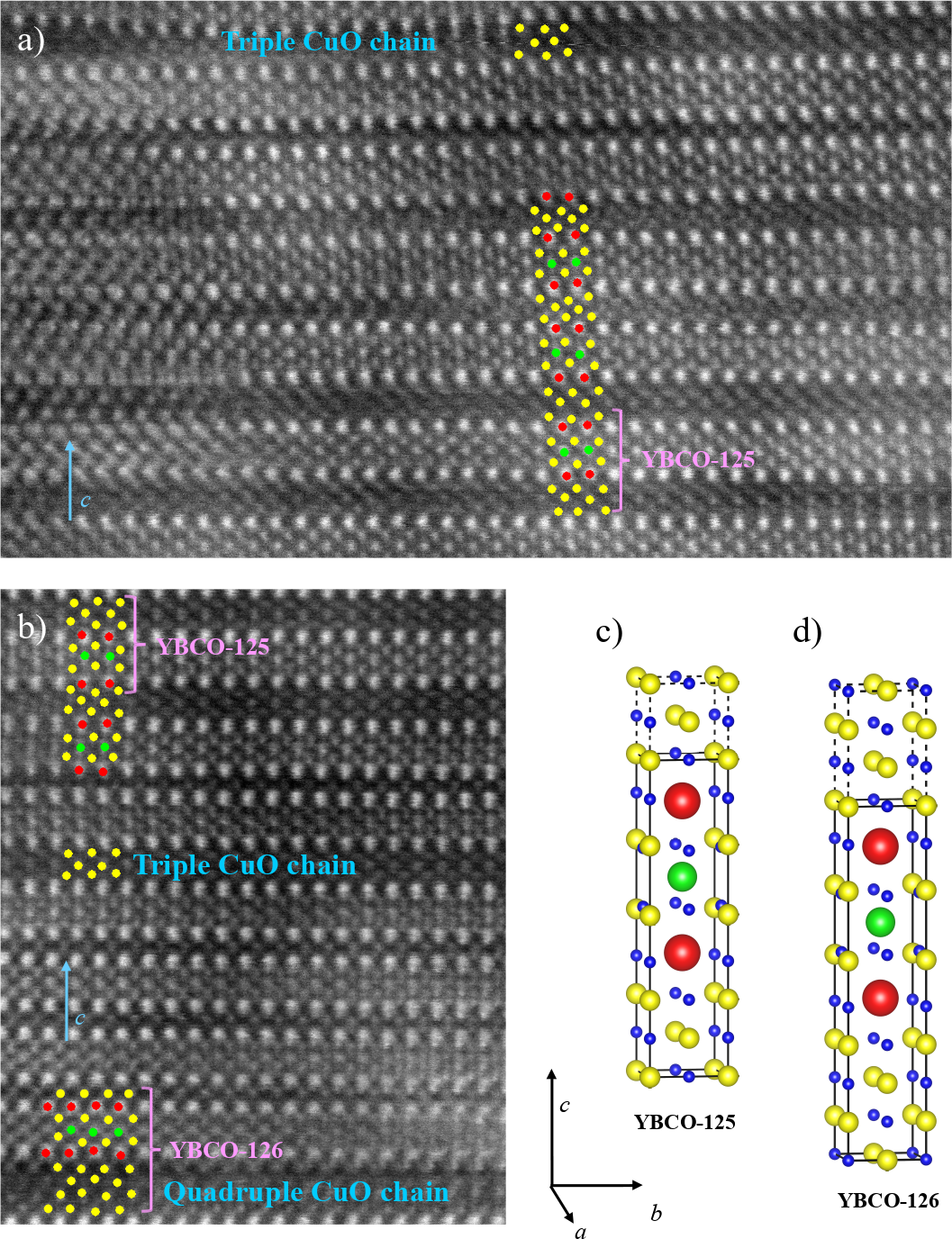}
\caption{Evidence for phase transformation to YBCO-125 and YBCO-126 in HP-annealed films. The high-resolution STEM image in panel (a) shows regions with three CuO chains per unit cell.  The image in panel (b) shows regions with four CuO chains per unit cell.  The lattice structures of YBCO-125 and YBCO-126 are shown in (c) and (d) respectively, with Y, Ba, Cu and O color-labelled as green, red, yellow and blue.}
\end{figure}

In some of our HP-annealed films, regions with triple-CuO and quadruple-CuO chains were observed, as shown by the high-resolution STEM image in Figure 5.  These regions constitute exotic Y-Ba-Cu-O phases with three and four CuO chains per unit cell, corresponding to YBCO-125 and YBCO-126 respectively, as illustrated in Fig. 5(c) and 5(d) by the lattice structures drawn with VESTA software \cite{vesta}. YBCO-125 and YBCO-126 have been seen as defect phases in trace amounts before \cite{ramesh3,senaris,hashimoto,fu}, but not yet been systematically synthesized despite attempts in up to 3000 bar of $\rm{O_2}$ \cite{karpinski2}. Evidently, the unique thermodynamics of thin-film samples facilitates the formation of Y-Ba-Cu-O phases that do not form in bulk samples. The phase conversion from YBCO-123 to YBCO-125 and to YBCO-126 can be represented by the following equations:

\begin{equation*}
\begin{split}
  \rm{YBa_2Cu_3O_{7-\delta} + 2CuO \xrightarrow{HP\ O_2} YBa_2Cu_5O_{9-\delta}}\\ 
  \rm{YBa_2Cu_3O_{7-\delta} + 3CuO \xrightarrow{HP\ O_2} YBa_2Cu_6O_{10-\delta}}
\end{split}
\end{equation*}

The multiple CuO chains in the YBCO-125 and YBCO-126 phases offer new opportunities for investigating the rich physics of the CuO chains in Y-Ba-Cu-O. First, a comparative study of these multi-chain phases will shed further light on the proximity-induced pairing in the CuO chains \cite{NgaiPRL}, and also help to clarify whether the quasi-1D ribbons formed by these chains can sustain pairing on their own \cite{kuwabara, Berg, matsukawa}.  Second, the multiple CuO chains may be used to tune the coupling strength between the CuO$_2$ planes \cite{stern1995}, which is known to affect the $T_c$ in cuprates \cite{locquet, sato, pavarini, Chakra, hirata}. Thus, successful isolation of these exotic Y-Ba-Cu-O phases will allow us to explore novel superconductivity in the CuO chains and potentially also to enhance $T_c$ \cite{cavalleri}. 

At present, the YBCO-125 and YBCO-126 phases in our thin-film samples are not yet robust enough for their $T_c$ to be determined resistively.  To better isolate these phases, our superoxygenation technique will need to be refined.  Some of the refinement strategies we are pursuing include: 1) varying the annealing pressure and temperature; 2) incorporating excess CuO directly into the PLD target; 3) varying the film thickness; and 4) using alternative substrates with different lattice parameters to tune the heteroepitaxial strain. The first two strategies are guided by thermodynamic and kinetic considerations, towards finding higher-purity regions in an expanded phase diagram. The latter two strategies are motivated by our STEM observation that more phase conversion tends to occur closer to the film surface, which suggests that ionic diffusivity and lattice strain \cite{zhang} may also affect the phase stability.

\section{IV. CONCLUSION}
In summary, we have demonstrated that high-oxidation phases of Y-Ba-Cu-O can be synthesized in epitaxial thin-film samples via superoxygenation.  The YBCO-123 thin films were grown by PLD on LSAT substrate, and then annealed at up to 700 atm O$_2$ and 900$^\circ$C in conjunction with Cu enrichment by solid-state diffusion.  STEM revealed clear formation of YBCO-247 and YBCO-248 phases in the post-annealed films, as well as regions of YBCO-125 and YBCO-126 phases having triple-CuO and quadruple-CuO chains. XRD and XAS measurements showed that higher Cu enrichment resulted in greater phase conversion to higher-oxidation phases, and that similarly annealed YBCO-123 powders do not undergo phase conversion. As proof of concept, our results open a novel route of synthesis towards discovering more complex phases of high-$T_c$ cuprates and other superconducting oxides.

\section{ACKNOWLEDGEMENTS}
This work is supported by NSERC, CFI-OIT and CIFAR. The electron microscopy work was carried out at the Canadian Centre for Electron Microscopy, a National Facility supported by the Canada Foundation for Innovation under the Major Science Initiative program, McMaster University and NSERC. The XAS work was performed at the Canadian Light Source, which is supported by NSERC, NRC, CIHR, and the University of Saskatchewan.

\end{document}